\documentclass[aps,twocolumn,preprintnumbers,%
superscriptaddress,floatfix]{revtex4}

\usepackage[dvips]{graphics}
\usepackage{amsmath,amsfonts,bm}
\usepackage{psfig}

\begin{document}
\newcommand{\be}{\begin{eqnarray}}
\newcommand{\ee}{\end{eqnarray}}
\def\lsim{\mathrel{\rlap{\lower3pt\hbox{\hskip1pt$\sim$}}
     \raise1pt\hbox{$<$}}} %less than or approx. symbol
\def\gsim{\mathrel{\rlap{\lower3pt\hbox{\hskip1pt$\sim$}}
     \raise1pt\hbox{$>$}}} %greater than or approx. symbol
\def\N{${\cal N}\,\,$}
\newcommand\<{\langle}
\renewcommand\>{\rangle}
\renewcommand\d{\partial}
\newcommand\LambdaQCD{\Lambda_{\textrm{QCD}}}
\newcommand\tr{\mathrm{Tr}\,}
\newcommand\+{\dagger}
\newcommand\g{g_5}

\title{Study of Strangeness Condensation by Expanding About
the Fixed Point of the Harada-Yamawaki Vector Manifestation}
\author {G.E. Brown}
\affiliation { Department of Physics and Astronomy\\ State
University of New York, Stony Brook, NY 11794-3800, U.S.A.}
\author{Chang-Hwan Lee}
\affiliation{Department of Physics, Pusan National University,
              Pusan 609-735 and
              \\
Asia Pacific Center for Theoretical Physics, POSTECH, Pohang
790-784, Korea}
\author{Hong-Jo Park}
\affiliation{Department of Physics, Pusan National University,
              Pusan 609-735, Korea }
\author{Mannque Rho}
\affiliation{ Service de Physique Th\'eorique,
 CEA Saclay, 91191 Gif-sur-Yvette c\'edex, France and\\
School of Physics, Seoul National University, Seoul 151-747,
Korea}
%\date{\today}
\begin{abstract}
Building on, and extending, the result of a higher-order in-medium
chiral perturbation theory combined with renormalization group
arguments and a variety of observations of the vector
manifestation of Harada-Yamawaki hidden local symmetry theory, we
obtain a surprisingly simple description of kaon condensation by
fluctuating around the ``vector manifestation (VM)" fixed point
identified to be the chiral restoration point. Our development
establishes that strangeness condensation takes place at $\sim 3
n_0$ where $n_0$ is nuclear matter density. This result depends
only on the renoramlization-group (RG) behavior of the vector
interactions, other effects involved in fluctuating about the bare
vacuum in so many previous calculations being ``irrelevant" in the
RG about the fixed point. Our results have major effects on the
collapse of neutron stars into black holes.
\end{abstract}

%\date{\today}

\newcommand\sect[1]{\emph{#1}---}

\maketitle

\sect{Introduction\label{intro}} The calculation of strangeness
condensation in neutron stars has got into great complexity
because the necessary interactions, especially those of strange
hadrons, are only partially known. The calculation within a chiral
formulation of Thorsson, Prakash, and Lattimer \cite{TPL1994} gave
the condensation density $n_c\simeq 3.1 n_0$, where $n_0$ is
nuclear matter density. This calculation was later modified by
other authors by introduction of higher order chiral perturbation
expansion, strange hyperon condensates, the role of $\Lambda
(1405)$ important for threshold $K^-p$ processes, etc., in which
extrapolations from low densities to densities $\sim 3 n_0$ had to
be made.

In this note, based on qualitative understanding of the
condensation process extracted from chiral perturbation theory
combined with renormalization group analysis, we propose to study
s-wave kaon condensation starting from the vector manifestation
(VM) fixed point discovered by Harada and Yamawaki in the hidden
local symmetry approach to effective field theory of
QCD~\cite{HY:VM,HY:PR}.

A surprise in the next-to-next chiral perturbation calculation of
the kaon condensation involving in-medium two-loop graphs in
\cite{LBMR} was that certain two-body correlations via effective
four-fermi operators were found to be $essential$ for driving the
condensation but played a negligible role in locating the critical
density. For instance, attractive four-fermi interactions
involving $\Lambda (1405)$ which figures crucially in threshold
$K^- p$ interactions (see \cite{borasoyetal} for a recent review
with references), are $necessary$ for the condensation but the
critical point is highly insensitive to the strength of the $K^- N
\Lambda (1405)$ coupling constant: Varying the coupling constant
by two orders of magnitude changed the critical density by only a
few \%.

This surprising result can be understood in terms of
renormalization group flow arguments~\cite{rge-lrs}. In
\cite{rge-lrs}, generic boson condensation in dense matter with
conditions commensurate with kaon condensation in neutron star
matter (i.e., chiral symmetry broken by the s-quark mass and the
presence of electron chemical potential $\mu_e$) was studied with
a toy model using the renormalization group technique developed by
Shankar~\cite{shankar} for Landau Fermi liquid theory. The model
consisted of a $``relevant"$
%\footnote{Throughout this paper,
%``relevance" and ``irrelevance" in the renormalization-group sense
%will put under a quotation mark.}
mass term (quadratic in boson field) and an $``irrelevant"$
boson-baryon four-point interaction term. (From here on, we put
under quotation mark the terminology in the RG sense.)
%~\footnote{It should be
%stressed that the boson condensation process treated here involves
%a renormalization group flow which is different from that of
%Shankar's Fermi liquid. In the case we have, the ``mass" term is
%$``relevant"$ whereas in Landau Fermi liquid, the effective mass
%does not flow since the Fermi surface is fixed by fiat, that is,
%the effective mass is a fixed point quantity there.}
It was shown there (corresponding to the case of Fig.2(c) in
\cite{rge-lrs}) that while condensation was driven by a
``relevant" term as is the standard case for phase change in field
theory, the $direction$ of the flow as well as the $fate$ of the
state crucially depended upon whether the $leading$ ``irrelevant"
interaction is attractive or repulsive: Condensation is inevitable
if attractive and does not take place if repulsive. Thus, the
attractive sign of the ``irrelevant" interaction was indispensable
for the condensation, a feature that is unusual in condensed
matter physics where marginal terms figure initially. However once
one is near the critical point, the ``irrelevant" attractive
interaction term that $determined$ the flow direction plays a
minor role.

In terms of the chiral perturbation calculation of \cite{LBMR},
what we have here is a consequence of an intricate interplay
between the chiral limit at which the kaon is a Goldstone particle
and the heavy-quark limit at which the kaon is a ``heavy meson."
In the two limits, condensation does not take place. {\it Kaons
condense because the kaon is neither heavy nor
light}.\cite{rge-lrs}

What we learn from the above observations is that the relevant
degrees of freedom for kaon condensation are not necessarily those
manifest in free space where elementary kaon-nucleon interaction
takes place but it would be more astute to identify the degrees of
freedom relevant in the vicinity of the condensation point. In
addressing this issue, we will start with our principal assumption
that kaon condensation takes place in the vicinity of -- but below
-- $n_{\chi SR}$ at which the spontaneously broken chiral symmetry
in Goldstone mode is restored to Wigner mode. This assumption
allows us to consider fluctuating around $n_{\chi SR}$ rather than
around the free space $n=0$. The appropriate formalism for this is
Harada-Yamawaki hidden local symmetry theory~\cite{HY:PR}. Were it
not for the presence of electron Fermi sea, the flow would wind up
at the VM fixed point. However the decay of electrons into
kaons~\cite{BKRT} followed by the condensation of kaons at $n_c
\lsim n_{\chi SR}$ stops the HLS flow.
%---------------------------------------------------------

\sect{Physics near the VM fixed point} In applying
Harada-Yamawaki(HY) hidden local symmetry (HLS) theory to dense
(and hot) matter, the key ingredient is the flow of the principal
parameters of the HLS Lagrangian, the hidden gauge coupling $g$,
the ``bare" (parametric) pion decay constant $F_\pi$ and the ratio
$a\equiv (F_\sigma/F_\pi)^2$ (where $F_\sigma$ is the decay
constant of the would-be Goldstone scalar that gives rise to the
longitudinal component of the massive gauge field)~\cite{HY:PR}.
These parameters flow as the scale is varied in a variety of
different directions with a variety of fixed points in the
one-loop RGE. The most important point however is that the coupled
RGEs flow to one unique fixed point called vector manifestation
(VM) fixed point when the HLS theory is matched to QCD at a
suitable matching scale $\Lambda_M$. The VM fixed point turns out
to be
 \be
(\bar{g}, \bar{f}_\pi, \bar{a})= (0, 0, 1)
 \ee
where $f_\pi$ is the physical pion decay constant related to
$F_\pi$ with a quadratically divergent correction. HY show at
one-loop order that the VM fixed point is arrived at when the
quark condensate $\langle\bar{q}q\rangle$ -- the order parameter
of chiral symmetry -- goes to zero. This has been verified by
Harada and Sasaki~\cite{HS1} in heat bath as $T\rightarrow T_{\chi
SR}$ and by Harada, Kim and Rho\cite{HKR} in dense matter as
$n\rightarrow n_{\chi SR}$. We shall refer to the HLS theory
endowed with the vector manifestation as HLS/VM.

The most important physical implication of HLS/VM in hot/dense
matter is that the vector meson mass and the gauge coupling
constant go to zero as the critical point is approached in
agreement with the BR scaling~\cite{BR91}. Another implication
which has been less extensively exploited up to date is that the
constant $a$ goes to 1 at the VM. Most relevant to our present
work is that the HLS/VM theory simplifies the calculation
enormously near the VM point because of the particularly simple
limiting behavior of the constants. Now the question is: Can one
start from this point, instead of from the vacuum as is commonly
done, to do physics taking place not on top of but near the VM
point? This issue was addressed in \cite{MR-nagoya} where several
non-trivial and interesting cases were given; i.e.
% \begin{itemize}
% \item
(1) The chiral doubling of heavy-light hadrons predicted more
 than a decade ago using sigma model built on the (matter-free)
 vacuum~\cite{chiraldoubling} and confirmed by the BaBar
 and CLEO collaborations in $D$ mesons can be very simply
 understood in terms of HLS/VM in the light-quark sector starting
 from the VM fixed point~\cite{HRS}. Here $a$ remains equal to
 1 to the order considered; (2)
 %\item
 the $\pi^+$-$\pi^0$ mass splitting can be very well
 reproduced in HLS theory with $a=1$ and $g\neq
 0$~\cite{pionmass}; (3)
 %~\footnote{That
 %$a=1$ is of the special importance here in connection with the
 %notion of ``little Higgs" and ``theory space locality" in the
 %Standard Model. The emergence of hidden local symmetry and the
 %notion of theory space locality are relevant for extending
 %low-energy effective field theory to the regime where the effective theory
 %breaks down and should be ``ultraviolet completed" to a fundamental
 %theory. See  \cite{pionmass}.}
 %\item
 in the presence of baryonic matter, even when one is far
 away from the VM fixed point, $a$ flows precociously to 1. For
 instance, $a\approx 1$ in the EM form factor for the nucleon as
 discussed in \cite{BR04}; (4)
 %\item
 in the presence of temperature, $a$ flows to 1 and vector
 dominance in the EM form factor of the pion breaks down maximally
 at chiral restoration~\cite{HS}.
 This vector dominance violation has an important
 implication in dilepton production in heavy-ion
 collisions~\cite{na60}.
 %\end{itemize}
%----------------------------------------------------

\sect{Kaon Fluctuation Around the VM Fixed Point} A clear hint
from the above discussions is that in the presence of medium,
temperature and/or density, $a$ tends to go to 1 quickly whether
or not one is near the VM fixed point with $g=0$. Since kaon
condensation occurs in dense medium and most likely in the
vicinity of the VM fixed point, it should be much more profitable
to expand around the VM fixed point with $a=1$ than around the
matter-free vacuum. The strategy is quite analogous to that used
in \cite{HRS} for the chiral doubler splitting of heavy-light
hadrons. One starts with a bare Lagrangian fixed by Wilsonian
matching to QCD at the matching scale $\Lambda_M$.

Now near the VM fixed point, the quark condensates are rotated
out, so the sigma term which is more ``irrelevant" than the
Weinberg-Tomozawa term will no longer be important. Furthermore,
the similarly ``irrelevant" four-point interactions that intervene
to drive kaon condensation when initiated from the matter-free
vacuum such as those involving $\Lambda (1405)$ and p-wave
interactions involving hyperons would have flowed to zero at the
VM fixed point as mentioned above. What is left would then be the
Weinberg-Tomozawa-type term -- which is the {\it least
``irrelevant"} from the point of view of RGE -- from the exchange
of the $\omega$-meson (and the $\rho$-meson in non-symmetric
matter) between the kaon and the baryon.

Now to compute the Weinberg-Tomozawa term in dense medium, we need
baryons in the theory. But Harada-Yamawaki HLS/VM theory has no
baryons. Baryons must therefore be generated as skyrmions. Near
the VM fixed point, however, baryons are not the correct fermion
degrees of freedom. We postulate that the relevant fermionic
degrees of freedom at some density above that of nuclear matter -
not yet precisely pinned down -- in HLS/VM theory are constituent
quarks or quasiquarks~\cite{HKR}, just as in heat bath above the
``flash temperature" $T_{flash}\sim 125$ MeV indicated in lattice
calculations. The flash temperature occurs at the point nucleons
change into quasiquarks~\cite{BLR-star}. Its value of 125 MeV is
obtained by lattice calculations. (See Fig.1 and Eq. (3) of
\cite{BLR-slavery}).

We assume that the quasiquark notion is applicable for $n\gsim
n_0$. The density $n_0$ may be regarded as an analog to the flash
temperature $T_{flash}$~\cite{BLR-star} in the sense that the
gauge coupling $g$ starts dropping and $a\simeq 1$ from that point
to the critical point. This will become clearer later, where the
results are not very sensitive to the choice of the change-over
density which could be $\sim 2 n_0$ without changing much our
scenario.
%Basically what we are saying is that at some density
%nucleons loosen -- in some mechanism that is not yet precisely
%understood -- into lightly bound quasiquarks and then these, as in
%Nambu-Jona-Lasinio, make a transition into current quarks at the
%chiral restoration density $n_{\chi SR}$.}
In the mean field, we
estimate that the $\omega$ exchange
 \be V_{K^-}(\omega) =
-\frac{3}{8 F_\pi^2} n
 \ee will give a
potential \be V_{K^-}(\omega) \simeq -57\ {\rm MeV} \;
\frac{n}{n_0}\label{estimate}
 \ee when using the parametric pion decay constant\cite{WaasWeise}
  $F_\pi \simeq 90\
{\rm MeV}$. Note that $F_\pi$ should be distinguished from the
physical pion decay constant $f_\pi$, the order parameter that
goes to zero whereas the former does not.

Now the Walecka vector mean field, because of the three
quasiquarks in the nucleon as compared with one nonstrange
anti(quasi)quark in the $K^-$, should be
 \be V_N(\omega) = -3 V_{K^-} (\omega)\ .
 \ee
But at $n=n_0$, $V_N(\omega)\simeq 171$ MeV is well below the
empirical value of more like $V_N (\omega) \sim 270$ MeV usually
used. Clearly we are missing something here. What we are missing
is precisely the BR scaling of the pion decay constant.

{}From the analysis of deeply bound pionic atoms \cite{Suzuki2004}
we have learned that {\it in medium} the parametric $F_\pi$ must
be decreased
 \be F_\pi \rightarrow f_\pi^\star \approx 0.8 F_\pi
 \ee
$\sim 20 \%$ at $n=n_0$ in movement towards chiral
restoration.
%\footnote{At the mean field order we are considering,
%this quantity is indeed the order parameter, so we can say it is
%movement towards chiral restoration.}
The Walecka vector mean
field, obtained for $n=n_0$, already has this medium dependence
empirically built into it; i.e., using $f_\pi^\star$ instead of
$F_\pi$ valid at tree order it is increased by $(1/0.8)^2$.
%As
%discussed above, with the Harada-Yamawaki vector manifestation
%\cite{HY:PR}, the physics can be better pinned down if we expand
%about the fixed point in this manifestation where the vector meson
%mass and coupling constants $m_V^\star$ and $g_V^\star$ go to
%zero.
%The fixed point is at the chiral restoration density
%$n_{\chi SR}$.

Our next task is to show that the critical density for strangeness
condensation is close to the density at the fixed point. For this,
most important for the extrapolation from fixed point to critical
density $n_c$ for kaon condensation is the renormalization group
flow in the parameter $a=(F_\sigma/F_\pi)^2$.

Using full (unquenched) lattice calculation results for
SU(2)$\times$SU(2), Park et al. \cite{PLB} have shown that the
degrees of freedom giving the entropy at $T_c$ in the finite
temperature chiral restoration transition
%identified with the
%Harada and Yamawaki fixed point
are the quasiquark-antiquasiquark bound states $\pi, \sigma$ and
vector and axial vector mesons that go massless at $T_c$. It seems
reasonable to expect that these are the only degrees of freedom
left $also$ at $n_{\chi SR}$.
%Note that the chiral condensate $\langle\bar{q}
%q\rangle$ which multiplies the strangeness sigma term
%$\Sigma_{KN}$ in calculations of strangeness condensation about
%zero density is zero at $n_{\chi SR}$.
%--------------------
%---------------------------------------------------

\sect{Behavior of Parameters in the Vicinity of the Fixed Point}
First we should estimate (in the chiral limit) at which density
$n_{\chi SR}$ the fixed point is reached. We do this by finding
out at which density $m_\rho^\star$ goes to zero.

Although initially $m_\rho^\star$ decreases as $\sqrt{\langle \bar
q q\rangle^\star}$ following the scaling of $F_\pi^\star$
%(see argument above following Eq.~(\ref{eq19}))
the Harada and Yamawaki work shows that once it starts dropping it
scales as $\langle \bar q q\rangle^\star$. (See the empirical
verification of this in Koch and Brown \cite{KochBrown93} who
showed that the entropy matched that in LGS if the meson masses
were allowed to scale as $\langle\bar q q\rangle^\star$, which was
referred to as ``Nambu scaling.") As noted by Brown and Rho
\cite{BR04}, $g$ does not seem to scale up to nuclear matter
density $n_0$, but then Nambu scaling sets in. Nambu scaling is
$\sqrt 2$ times faster than the initial scaling of $m_\rho^\star$
from $n=0$ to $\sim n_0$, which decreases $m_\rho^\star$ by 20\%.
Thus, we believe in the interval $n_0$ to $2 n_0$, $m_\rho^\star$
will decrease $\sqrt 2$ times 20\%, or $\sim 28\%$, and the same
from $2 n_0$ to $3 n_0$, and from $3 n_0$ to nearly $4 n_0$ where
$m_\rho^\star=0$ in the chiral limit. Thus, the fixed point at
$n_{\chi SR}$ is at $n\lsim 4 n_0$.

{}From the Brown and Rho \cite{BR04} argument that $g^\star$
scales as $m_\rho^\star$ for $n > n_0$, but up to $n_0$, $g$
remains constant, whereas $m_\rho^\star$ scales, we find
 \be
\frac{{g^\star}^2}{{m_\rho^\star}^2} = \frac{1}{a^\star
{F_\pi^\star}^2}\approx \frac{1}{a^\star} \left( \frac{1}{0.8
F_\pi}\right)^2
 \ee
and  at $n_{\chi SR}$, $a^\star=1$, together with $m_\rho^\star=0$
and $g^\star=0$. Compared with the matter-free expression
$m_\rho^2=2 F_\pi^2 g^2$, which is the KSRF relation, we see that
 \be
\frac{[{g^\star}^2/{m_\rho^\star}^2]_{\rm fixed \; point}}
     {[{g}^2/{m_\rho}^2]_{\rm zero \; density}}
     = \frac{[a F_\pi^2]_{\rm zero \; density}}
            {[a^\star {F_\pi^\star}^2]_{\rm fixed \; point}}
     \simeq\frac{2}{0.8^2} \simeq 3.1.\label{factor}
 \ee
Thus, the mean field felt by the $K^-$ is increased by the factor
3.1 when the scaling of both $F_\pi^\star$ and $a^\star$ are
included, at the fixed point at $n_{\chi SR}$, the final doubling
coming from the scaling in $a^\star$. Note that the scalar
contribution from the rotation of the sigma term is gone.

Now we calculate $m_{K^-}^\star$ for neutron rich matter. If we
calculate for 90\% neutrons and 10\% protons at $n_c=3.1 n_0$
without medium dependence
 \be V_{K^-} = %- \frac{1}{9}\frac{g_\omega^2}{m_\omega^2}
-\frac{1}{a F_\pi^2} \left(\frac{x_n}{2}+x_p\right) n_c = - 129\
{\rm MeV}
 \ee
where we have included $\rho$ as well as $\omega$ exchange, and
$x_{n,p}$ are the neutron and proton fractions.
But now we have to incorporate the enhancement factor
(\ref{factor}). Thus, taking $n=4 n_0$ and multiplying by 3.1 in
order to take into account the medium dependence, we have
 \be V_{K^-} \approx
-\frac{4}{3.1} \times 3.1 \times 129 \ {\rm MeV} = -516\ {\rm MeV}
\lsim - m_{K^-}.
 \ee
Thus, the vector mean field at $n=4 n_0$ is sufficient to bring
the $m_{K^-}^\star$ to zero. That is, the vector mean field is
strong enough to bring $m_{K^-}^\star$ to zero at $n_{\chi SR}$,
which would imply $\mu_e=0$ for strangeness condensation at this
density. Of course $\mu_e \neq 0$, so the condensation must take
place at a lower density. But the fact that  $m_{K^-}^\star$ goes
to zero at $n_{\chi SR}$ means that the $K^-$ behaves more like a
normal nonstrange meson than a Goldstone boson. This aspect may be
related to the RGE finding mentioned above that {\it kaon
condensation takes place because the kaon is neither heavy nor
light}.

In moving to $n\simeq 3 n_0$, the change in density will produce
$m_{K^-}^\star (\frac 34 n_{\chi SR}) \simeq m_{K^-}/4 \simeq 124
$ MeV. Furthermore, the renormalization group parameter $a$ may
have increased from the fixed point value of 1. Since $a$ tends
rapidly to 1 in the presence of baryonic matter~\cite{MR-nagoya},
we do not expect that it will be much different from 1 in the
density regime we are dealing with. An upper limit may be taken to
be $a\simeq 4/3$, the value at the matching scale $\Lambda_M\sim
1.1$ GeV which corresponds to the large $N_c$ limit of
Harada-Yamawaki's HLS bare Lagrangian\cite{HY:PR} which is also
the value obtained in holographic dual QCD that exploits AdS/QCD
in string theory\cite{sugimoto}. Now this increases
$m_{K^-}^\star$ by that factor so that
 \be m_{K^-}^\star\simeq 124\times 4/3\ {\rm MeV} \simeq  165\ {\rm MeV}
 \ee
which should equal ${\mu_e}$. It is somewhat smaller than, but not
too far from, the Thorsson et al. $\sim 220$ MeV \cite{TPL1994}.

%--------------------------------------
\sect{Special Properties of the Vector Mesons in the Hadron Free
Zone} Harada and Yamawaki \cite{HY:PR} show that $m_\rho^\star$
goes to zero at the fixed point, at the same rate as $g^\star$, so
that $g^\star/m_\rho^\star$ goes to a constant. This was observed
even earlier \cite{BR96} in studying the lattice calculations of
the quark number susceptibility, and discussed in \cite{BR04}. The
isoscalar quark number susceptibility is the same as the isovector
one. Three of the former can be added together to give the
behavior of the $\omega$, whereas two of the isovector quark
susceptibilities cancel each other to make up the $\rho$, giving
 \be g_\omega = 3 g_\rho
 \ee
with $g_\rho=\frac 12 ag$. As $T\rightarrow T_{\chi
SR}$~\cite{HS}, and $n\rightarrow n_{\chi SR}$~\cite{HKR}, HLS/VM
predicts that the interactions between hadrons go to zero; i.e.,
the {\it hadronically free region} is reached. This phenomenon was
referred to as {\it hadronic freedom} in \cite{BLR-slavery}. From
lattice calculations and the STAR experiments~\cite{STAR:04}, we
can reconstruct how this happens in hot matter in some
detail~\cite{BLR-star} and then conjecture what could happen in
dense matter.

The reasoning goes as follows.

Baryonic and hyperonic interactions would not be expected to
survive the hadronic freedom region. We know how this comes about
in temperature~\cite{BLR-slavery} from lattice calculations: The
nucleon breaks up into three loosely bound constituent quarks at
$T\sim 125$ MeV and then they lose their masses, becoming
noninteracting current quarks as $T\rightarrow T_{\chi SR}=175$
MeV. This means that the baryonic interactions go to zero as
$T\rightarrow T_{\chi SR}$. In the same vein, we predict that the
same phenomenon takes place as $g^\star\rightarrow 0$ at the fixed
point in finite density. However that is the region of density
where kaons will condense. The matter is supported by electron
pressure and flows towards the VM fixed point \`a la HLS/VM until
it ``crashes" with the kaons condensing when electrons decay into
kaons. Then suddenly a neutron star will fall into a black hole in
a light-crossing time and never be seen again. This scenario will
be developed in more detail in a separate
publication~\cite{BL-tocome}.
%%%%%%%%%%%%%%%%%%%%%%%%%%%%%%%%%%%%%%%%%%%%%%%%%%%%%%%%%%%%%%%%%

\sect{Acknowledgments} GEB was supported in part by the US
Department of Energy under Grant No. DE-FG02-88ER40388. CHL was
supported by grant No. R01-2005-000-10334-0(2005) from the Basic
Research Program of the Korea Science \& Engineering Foundation.

%%%%%%%%%%%%%%%%%%%%%%%%%%%%%%%%%%%%%%%%%%%%%%%%%%%%%%%%%%%%%%%%%%%%%%%%%%%%%%%%

\end{document}